\documentclass[twocolumn,pra,aps,superscriptaddress,showpacs]{revtex4-1}

\usepackage[utf8]{inputenc}
\usepackage{booktabs}
\usepackage{mathtools}
\usepackage{amsmath,amssymb}
\usepackage{graphicx}
\usepackage{color}
\usepackage{appendix}

\definecolor{mygrey}{gray}{0.35}
\definecolor{myblue}{rgb}{0.2,0.2,0.8}
\definecolor{myzard}{cmyk}{0,0,0.05,0}
\definecolor{mywhite}{rgb}{1,1,1}
\definecolor{mywhite}{rgb}{1,1,1}
\definecolor{myred}{rgb}{1,0.,0.3}

\usepackage{hyperref}

\newcommand{\bra}[1]{\left\langle #1\right|}
\newcommand{\ket}[1]{\left| #1\right\rangle}

\newcommand\cc{\mathbf{c}}
\newcommand\dd{\mathbf{d}}

\newcommand\nn{\mathbf{n}}

\newcommand\kk{\mathbf{k}}
\newcommand\qq{\mathbf{q}}

\newcommand\bath{\mathrm{bath}}

\newcommand\KK{\mathrm{K}}
\newcommand\hc{\mathrm{H.c.}}
\newcommand\vac{\mathrm{vac}}
\newcommand\intt{\mathrm{int}}

\newcommand\BC{\mathrm{BC}}
\newcommand\BS{\mathrm{BS}}

\begin{document}

\title{Anisotropic quantum emitter interactions in two-dimensional photonic-crystal baths}

\author{A.~Gonz\'{a}lez-Tudela}
\affiliation{Instituto de F\'isica Fundamental IFF-CSIC, Calle Serrano 113b, Madrid 28006, Spain.}
\email{a.gonzalez.tudela@csic.es}
\affiliation{Max-Planck-Institut f\"{u}r Quantenoptik Hans-Kopfermann-Str.~1. 85748 Garching, Germany.}
\author{F.~Galve}
\affiliation{I3M (UPV-CSIC) Institute for Instrumentation in Molecular Imaging, Universidad Polit\'ecnica de Valencia, 46022, Spain}
\email{fernando.galve@i3m.upv.es}

\begin{abstract}
Quantum emitters interacting with two-dimensional photonic-crystal baths experience strong and anisotropic collective dissipation when they are spectrally tuned to 2D Van-Hove singularities. In this work, we show how to turn this dissipation into coherent dipole-dipole interactions with tuneable range by breaking the lattice degeneracy at the Van-Hove point with a superlattice geometry. Using a coupled-mode description, we show that the origin of these interactions stems from the emergence of a qubit-photon bound state which inherits the anisotropic properties of the original dissipation, and whose spatial decay can be tuned via the superlattice parameters or the detuning of the optical transition respect to the band-edges. Within that picture, we also calculate the emitter induced dynamics in an exact manner, bounding the parameter regimes where the dynamics lies within a Markovian description. As an application, we develop a four-qubit entanglement protocol exploiting the shape of the interactions. Finally, we provide a proof-of-principle example of a photonic crystal where such interactions can be obtained.
\end{abstract}

\maketitle

\section{Introduction \label{sec:intro}}

Engineering long-range dipole-dipole interactions is a new frontier in atomic and condensed matter physics. These interactions lead to a variety of exotic phenomena as compared to those appearing in systems with short-range interactions, such as non-local transmission of correlations~\cite{hauke13a,richerme14a,maghrebi16a} or fast equilibration~\cite{kastner11a,vodola14a,eldredge17a}. Moreover, they also yield long-range entanglement~\cite{shahmoon13a,shahmoon16a}, non-trivial self-organization patterns~\cite{chang13a,eldredge16a}, and can be used for quantum simulation~\cite{cirac12a} of chemistry~\cite{luengo18a} or other many-body problems~\cite{bernien17a}. Current experimental implementations to obtain these interactions are based on dipolar~\cite{trefzger11a} or Rydberg gases~\cite{lukin03a,saffman10a} and cavity QED~\cite{ritsch13a}. However, they offer limited shape tunability and, some of them, are unavoidably accompanied by dissipation. A timely alternative is based on photonic bandgaps appearing in photonic crystals~\cite{joannopoulos_book95a,chang18a}. When quantum emitters (QEs) are spectrally tuned to a photonic band-gap, the associated photonic crystal dissipation vanishes, and photons localize around them [forming the so-called qubit-photon bound states (BSs)~\cite{bykov75a,john90a,kurizki90a,tanaka06a,calajo16a,shi16a}] which can mediate tunable and long-range dipole-dipole interactions between the QEs~\cite{devega08a,navarretebenlloch11a,douglas15a,gonzaleztudela15c}. The spatial dependence of these BSs (and interactions), provided by the band-edge behaviour and dimensionality of the photon field, has been typically considered to be isotropic in the literature~\cite{devega08a,navarretebenlloch11a,douglas15a,gonzaleztudela15c}. It is therefore timely to consider further ways of shaping these interactions and expand the toolbox of photonic-crystal mediated interactions.

In two dimensional photonic crystals, the interplay between dimensionality and energy dispersion leads to strong features in the QE emission profiles, e.g., directional emission in Van Hove singularities~\cite{langley96a,mekis99a}. However, these features generally appear at spectral regions within the photonic bands, leading to strong collective effects but in the dissipative scenario~\cite{galve17a,gonzaleztudela17a,gonzaleztudela17b,galve18a}. In this work, we show how to transform this dissipation into anisotropic dipole-dipole coherent interactions by introducing an extra periodicity in the photonic crystal structure (superlattice). We focus on how to do it for 2D Van-Hove singularities in square geometries~\cite{galve17a,gonzaleztudela17a,gonzaleztudela17b,galve18a}, but our conclusions extend to other 2D geometries~\cite{galve18a,gonzaleztudela18c} or higher dimensions~\cite{gonzaleztudela18d}. 

The manuscript is structured as follows: in Section~\ref{sec:setup} we describe the system we study, focusing on the bath properties. In Section~\ref{sec:single} we calculate the exact dynamics for a single QE interacting with such photonic baths, characterize the regions of Markovian/Non-Markovian behaviour of the dynamics, and, more importantly, show that an extra photon BS appears inheriting the anisotropic properties of the directional emission at the Van-Hove point. In Section~\ref{sec:many}, we explore the interactions emerging from these BSs when many QEs are coupled to the bath, and exploit them to design a protocol to entangle four disconnected qubits through an auxiliary one. Afterwards, in Section~\ref{sec:bath} we show a proof-of-principle photonic crystal implementation that fulfills the required properties, and summarize our findings in Section~\ref{sec:conclu}.

\section{System \label{sec:setup}}

\begin{figure*}[tb]
\centering
\includegraphics[width=0.9\linewidth]{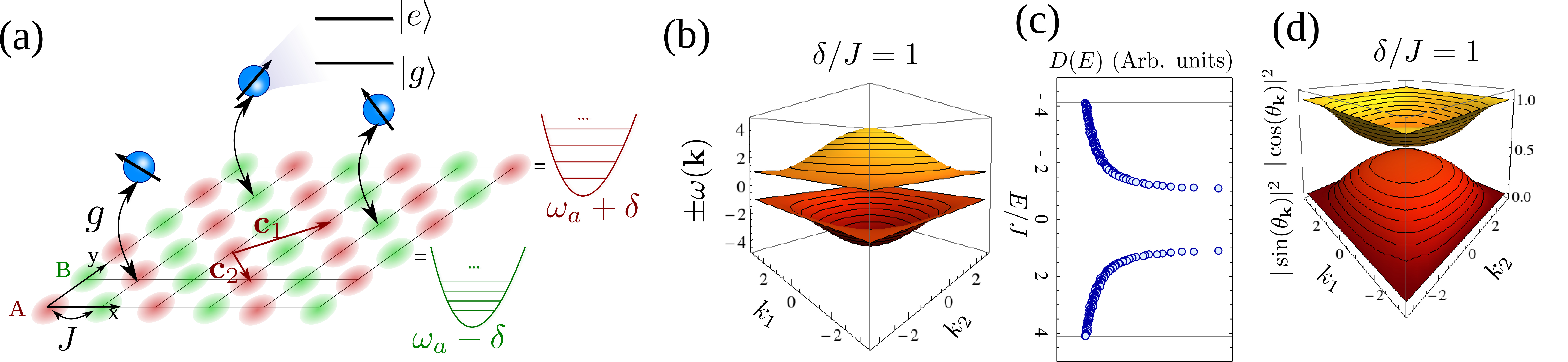}
\caption{(a) General setup: $N_e$ QEs are coupled with strength $g$ to a photonic bath with a superlattice geometry (two interspersed square lattices), with primitive vectors $\cc_{1,2}$, and alternating energy $\omega_a\pm \delta$. The bath excitations hop to their four nearest neighbours with the same rate $J$. (b) Upper/lower bath bands for an energy off-set $\delta=J$. (c) Numerical density of states calculated for the parameters of panel (b). (d) $|\cos(\theta_\kk)|^2$ and $|\sin(\theta_\kk)|^2$ as defined in Eq.~\ref{eq:cossin} for $\delta=J$.}
\label{fig:1}
\end{figure*}

The setup that we consider is sketched in Fig.~\ref{fig:1}: $N_e$ independent QEs are coupled through an optical transition, of frequency $\omega_e$, to a two-dimensional photon bath.  Their intrinsic Hamiltonian is given by (we take $\hbar=1$ for the rest of the manuscript): $H_S=\omega_e\sum_{j=1}^{N_e} \sigma_{ee}^j$, where we use the notation $\sigma_{\alpha\beta}^j=\ket{\alpha}_j\bra{\beta}$ for the spin operator of the $j$-th QE. For the bath we take the most simple model for a superlattice using a couple-mode description~\cite{haus91a}[see Fig.~\ref{fig:1}(a)]: it consists of two interspersed square lattices with $N\times N$ sites and primitive vectors $\cc_{1,2}=(1,\mp 1)$, where we take as unit of length the distance between the bosonic sites. The operators $a^\dagger_\nn,b^\dagger_\nn$ ($a_\nn,b_\nn$) are the creation (annihilation) operators of a bosonic excitation at the A/B sublattice  at site $\nn=(n_1,n_2)=\sum_{i=1}^2 n_i \cc_i$. We assume that the tunneling, at a rate $J$, only occurs between the four nearest neighbours at the A/B sublattices. Finally, to break the degeneracy between the 
A/B sublattices, we consider an energy off-set between the two sublattices such that $\omega_{a/b,\nn}\equiv \omega_{a}\pm \delta$. Thus, imposing periodic boundary conditions and defining $V^\dagger=(\hat{a}^\dagger_\kk,\hat{b}_\kk^\dagger)$, the bath Hamiltonian can be written in momentum space as $H_\bath=\sum_\kk V^\dagger H_\bath(\kk) V$, with:
\begin{align}
 \label{eq:HB}
 H_\bath(\kk)&=\left( \begin{array}{cc}
\omega_a+\delta& f(\kk) \\
f^*(\kk) & \omega_a-\delta
\end{array} \right)\,,
\end{align}
where $f(\kk)=J\left(1+e^{i k_1}+e^{i k_2}+e^{i (k_1+k_2)}\right)=|f(\kk)|e^{i\phi(\kk)}$, being $\kk=(k_1,k_2)$ the coordinates in the primitive vectors of the reciprocal space, i.e., $\dd_{1,2}=(\pi,\mp \pi)$. The operators $\hat{c}_\kk=\frac{1}{N}\sum_\nn e^{i \kk\cdot \nn} c_\nn$, with $c=a,b$ are the bath operators in $\kk$-space. Notice, we introduce the hat notation to distinguish the bath operators in real/momentum space. This Hamiltonian is diagonalizable by introducing a $\kk$-dependent transformation:
\begin{align}
\label{eq:Uk}
 U_\kk&=\left( \begin{array}{cc}
e^{i \phi(\kk)}\cos\theta_\kk & e^{i \phi(\kk)}\sin\theta_\kk \\
-\sin\theta_\kk & \cos\theta_\kk 
\end{array} \right)\,,
\end{align}
where:
\begin{align}
 &\cos\theta_\kk/\sin\theta_\kk=\pm\sqrt{\frac{\omega(\kk)\pm\delta }{2\omega(\kk)}}\,,\label{eq:cossin}\\
 &\omega(\kk)=\sqrt{|f(\kk)|^2+\delta^2}&\nonumber \\
 &=\sqrt{\delta^2 +16 J^2 \cos^2(\frac{k_1}{2})\cos^2(\frac{k_2}{2})}.
\end{align}

Using $U_\kk$, we diagonalize $H_\bath(\kk)=U_\kk^\dagger H_\bath(\kk) U_\kk=\sum_\kk \omega(\kk)\left(u_\kk^\dagger u_\kk-l_\kk^\dagger l_\kk\right)$, where $u_\kk/l_\kk$ denotes the eigenoperators for the upper/lower band of the photonic superlattice. Imposing $\delta=0$ the upper/lower bands touch, recovering the results of square geometries~\cite{galve17a,gonzaleztudela17a,gonzaleztudela17b,galve18a} where a 2D Van Hove singularity appears at $\omega_a$. When $\delta\neq 0$, the upper/lower bands extend between $[\omega_a-\sqrt{\delta^2+16J^2},\omega_a-\delta]$ and $[\omega_a+\delta,\omega_a+\sqrt{\delta^2+16J^2}]$, respectively, such that there exists a middle band-gap [see Fig.~\ref{fig:1}(b)]. Since this spectral region is the main focus of the article, we move to a rotating frame with $\omega_a$, such that $\omega_a\equiv 0$ in $H_B$, and  $\omega_e\rightarrow \Delta=\omega_e-\omega_a$ in $H_S$. Finally, we plot in Fig.~\ref{fig:1}(c) the numerical density of states $D(E)$ of this bath. As expected, $D(E)$ of both the upper/lower bands are symmetric with respect to $\omega_a$. At the upper/lower band-edges ($\pm \sqrt{16J^2+\delta^2}$), $D(E)$ has a discontinuous jump of the density of states typical of 2D isotropic band edges. On the contrary, the density of states diverges in the middle band-edges ($\pm \delta$). The $\kk$-points leading to these divergences, given by $k_1=\pm \pi$ and $k_2=\pm \pi$,  define a square in the reciprocal space which is the same than the one leading to directional emission in 2D Van-Hove singularities.

Now, let us finally write explicitly the coupling between QE and the bath modes, which reads:
\begin{align}
 H_\intt=\frac{1}{N}\sum_{c=a,b}\sum_{j=1}^{N_e}\left(g_{c,j} c^\dagger_{\nn_j} \sigma_{ge}^j+\hc\right)\,.
\end{align}
where we consider that the $j$-th QE only interacts locally with the A or B bath mode at site $\nn_j$, and with excitation-conserving terms. This Hamiltonian is a good approximation in the optical regime, where the typical frequencies are much larger than the coupling strength. When rewriting $H_\intt$, with the upper/lower band eigenoperators, we see that each QE couples the upper[lower] sublattice with $g_{u [l],j}(\kk)=e^{-i\left(\phi(\kk)+\kk\cdot\nn_j\right)}\cos(\theta_\kk)[\sin(\theta_\kk)]$ if the QE originally couples the $A$ sublattice, or $g_{[u] l,j}(\kk)=e^{-i\kk\cdot \nn_j}[-\sin(\theta_\kk)] \cos(\theta_\kk)$ if it couples to the B sublattice. In Fig.~\ref{fig:1}(d) we plot the square modulus of these functions and show that at the $\kk$-points which satisfy $\omega(\kk)\approx \delta$, $|\cos(\theta_\kk)/\sin(\theta_\kk)|\approx 1/0$. Thus, for fixed $\Delta$, the QE-bath dynamics is different depending on whether the QE is coupled to the A/B sublattice. This is qualitatively distinct from what happens in other 2D reservoirs considered in the literature~\cite{galve17a,gonzaleztudela17a,galve18a,gonzaleztudela17b,gonzaleztudela18c}, leading to new features in the dynamics.

\section{Single QE: anisotropic bound states \label{sec:single}}
\begin{figure}[tb]
\centering
\includegraphics[width=0.85\linewidth]{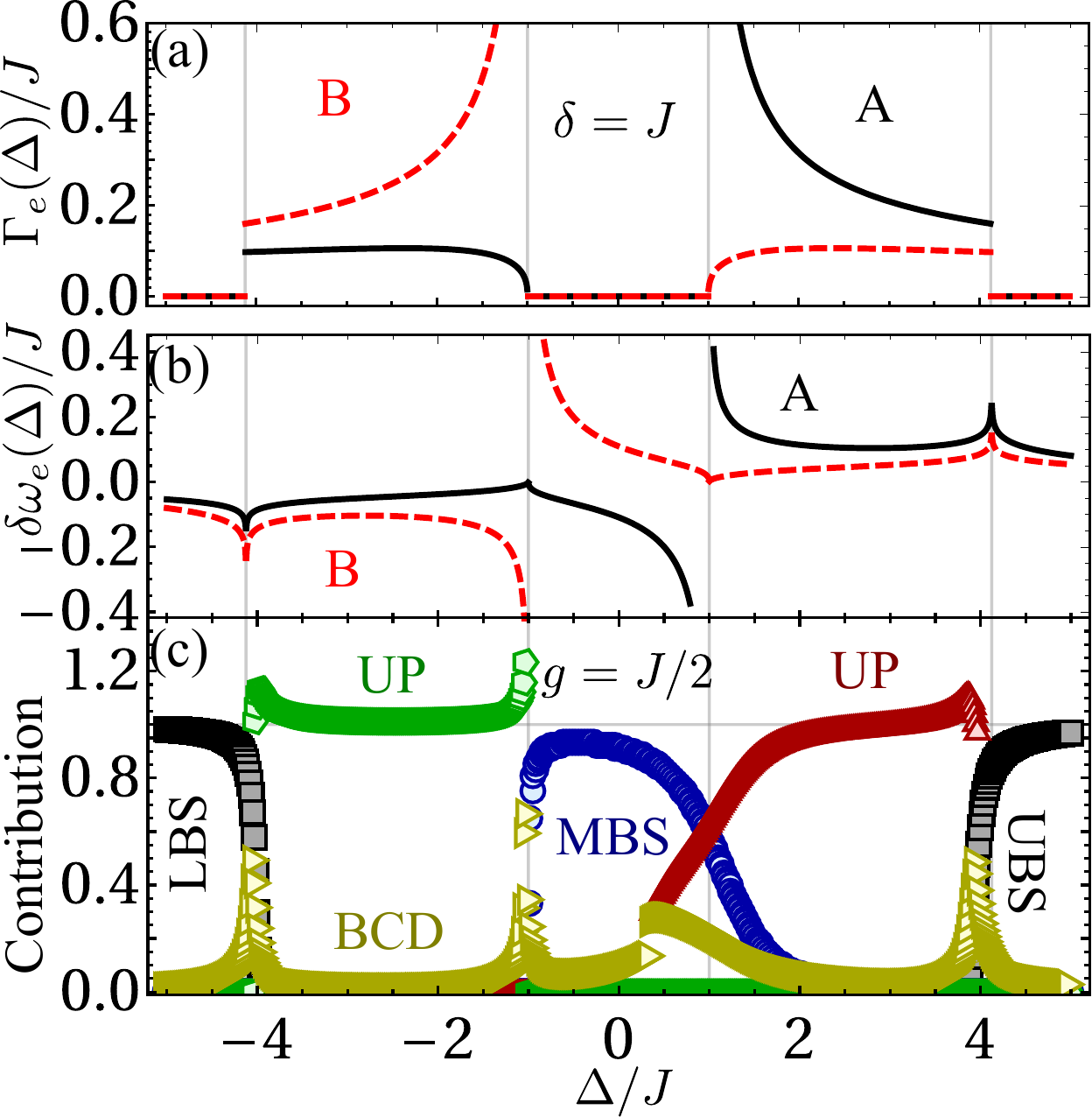}
\caption{(a)/(b) $\Gamma_e(E)/\delta\omega_e(E)$ for a QE coupled to a photonic superlattice bath with $\delta=J$. In solid black/dashed red line we plot the situation when the QE couples the A/B bath respectively. (c) Contributions to the dynamics of a QE at $t=0$ coupled with $g=J/2$ to the A sublattice of a bath with $\delta=J$. Legend.- black squares: upper/lower bound states (UBS); green pentagons/red triangles: unstable poles (UPs) from the upper/lower bands; yellow triangles: sum of all branch cut detours; blue spheres: middle bound states (MBS).}
\label{fig:2}
\end{figure}

First, let us characterize the dynamics of a single QE initially excited, i.e., $\ket{\Psi(0)}=\ket{e}\otimes\ket{\vac}_A\otimes\ket{\vac}_B$. The motivation is two-fold: first, we can solve it exactly using resolvent operator techniques~\cite{cohenbook92a,nakazato96a}, as explained in Refs.~\cite{gonzaleztudela17a,gonzaleztudela17b,gonzaleztudela18c}. Second, it helps us to identify the regions where the system admits a Markovian description, which we use in the second part of the manuscript. The key point is that probability to remain in the excited state $C_e(t)=\bra{\Psi(0)}e^{-i H t}\ket{\Psi(0)}$ can be written as:
\begin{align}
 C_e(t)=-\frac{1}{2\pi i}\int_{-\infty}^\infty  \frac{dE e^{-i E t}}{E+i0^+-\Delta-\Sigma_e(E+i0^+)}\,,\label{eq:ex}
\end{align}
where $\Sigma_e(z)$ is the QE self-energy which embeds the effect of the interaction with both the upper/lower baths~\cite{gonzaleztudela18c}.  In our case, $\Sigma_e(z)$ can be analytically calculated in the limit $N\rightarrow\infty$, yielding
\begin{align}
 \Sigma^{A/B}_{e}(z)=\frac{2 g^2 (z\pm \delta)}{\pi (z^2-\delta^2)}\KK\left[\frac{16 J^2}{z^2-\delta^2}\right]\,,\label{eq:analself}
\end{align}
where $\pm$ sign depends on whether the QE is coupled to the A/B lattices, and $\KK[m]$ the complete elliptic integral of the first kind~\cite{abramowitz66a}.

\emph{Perturbative treatment.} It consists of replacing $\Sigma_e(E+i0^+)\approx \Sigma_e(\Delta+i0^+)$ such that: $C_e(t)\approx e^{-i (\Delta+\Sigma_e(\Delta+i0^+)) t}$. Thus, the real/imaginary part of $\Sigma_e(E+i0^+)=\delta\omega_e(E)-i\Gamma_e(E)/2$ already provides us with the renormalization of the frequencies, $\delta\omega_e(E)$, and lifetimes, $\Gamma_e(E)$, of the QE excited state due to the interaction with the bath. This is what we plot in Figs.~\ref{fig:2}(a-b), where we observe several features which distinguish them from previous situations. First, we observe an asymmetry with respect to upper/lower bands in both $\delta\omega_e(E)/\Gamma_e(E)$. In particular, when the QE couples to the A/B sublattice the upper/lower edge of the middle band-gap shows a divergence, whereas in the other one the self-energy is strictly zero in spite of the divergence of the density of states in both edges [see Fig.~\ref{fig:1}(c)]. This asymmetry stems from the different $g_{u/l}(\kk)$ when the QE couples to the A/B modes, which cancels the divergence of the density of states in one of the band-edges. Since there is a clear symmetry $\Delta\rightarrow -\Delta$ and $A\rightarrow B$, from now on we focus on the situation when the QE couples to the A modes, and drop the superindex from $\Sigma_e(z)$. From Eq.~\ref{eq:analself}, one can prove [see Supp. Information] that the middle band-edges diverge as $1/\sqrt{x}$ [$\sqrt{x}$], as in 1D [3D isotropic] reservoirs. Both features have consequences in the emergence of an extra BS in the middle band-gap, as we show in the next Section.

\emph{Exact dynamics.}
 We already know from the literature that in the presence of bandgaps the QE dynamics may deviate substantially from the Markovian predictions [see Refs~\cite{john94a,tong10a,
longo10a,garmon13a,redchenko14a,lombardo14a,sanchezburillo17a}]. To fully characterize the different dynamical regimes, and bound the Markovian ones, we now integrate Eq.~\ref{eq:ex} exactly using complex analysis techniques [see Refs.~\cite{gonzaleztudela17b,gonzaleztudela18c,gonzaleztudela18d} and Supp. Information]. With  these methods, the dynamics of $C_e(t)$ can be split in several contributions: the one given by real poles of the integrand of Eq.~\ref{eq:ex}, whose origin is the emergence of a photon bound state (BS)~\cite{john90a,kurizki90a}; the one given by complex (unstable) poles (UPs) appearing in the analytical continuation of $\Sigma_e(z)$ to other Riemann surfaces; the extra contribution coming from the branch cut detours (BCDs) we have to define to apply Residue Theorem avoiding the non-analytical regions of the $\Sigma_e(z)$. In Fig.~\ref{fig:2}(c), we plot the absolute value of each contribution to $|C_e(0)|$ as a function of $\Delta$ for a QE coupled with $g=J/2$. This plot provides a very clear picture of the different dynamical regimes emerging on this setup and their origin. For example, when $|\Delta|\gg \sqrt{16J^2+\delta^2}$  the dynamics is dominated by the BS appearing in the upper/lower band-gaps. These are isotropic BSs like the ones appearing in other 2D reservoirs~\cite{gonzaleztudela15c,gonzaleztudela17b,gonzaleztudela18c}. When $\Delta$ lies within the upper/lower bands, the dynamics is dominated by the UPs, except for the points close to the edges, $\pm \delta,\pm\sqrt{16J^2+\delta^2}$, where the BCD contributes significantly. Apart from the asymmetry of the UP contribution between the two bands, a feature to highlight is that the UP of the upper band contributes to the dynamics even for $\Delta\in (-\delta,\delta)$, due to the divergent behaviour of $\delta\omega_e(\delta)$. However, the most relevant feature for this manuscript is the emergence of an extra bound-state in the middle band-gap, labeled as middle bound state (MBS) [in blue spheres in Fig.~\ref{fig:2}(c)]. Now, we focus on the MBS as it is qualitatively different from the BSs appearing in other 2D reservoirs~\cite{gonzaleztudela15c,gonzaleztudela17b,gonzaleztudela18c}.

\subsection{Middle bound state: existence and wavefunctions.}

\begin{figure}[tb]
	\centering
	\includegraphics[width=0.99\linewidth]{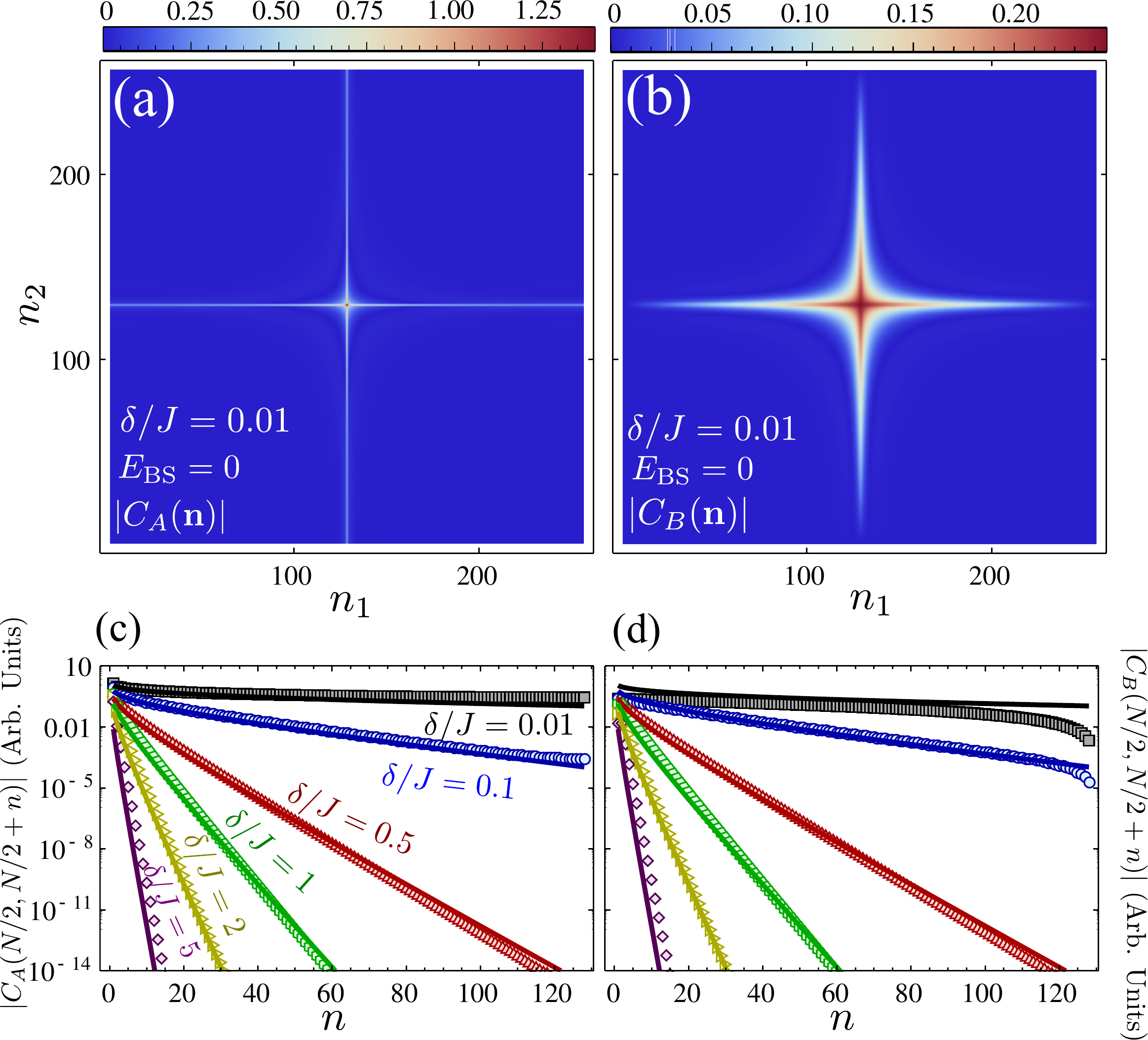}
	\caption{(a-b) $|C_{A/B}(\nn)|$ for $E_\BS=0$ of a QE coupled to an A mode. The bath parameters are $N=2^8$ and $\delta=0.01J$. (c-d) Horizontal cuts of the MBS wavefunction for the same lattice size than before, but several gaps $\delta/J$ as depicted in the legend. In solid markers we plot the numerical evaluation of the Fourier transform of Eqs.~\ref{eq:wavefunctionA}-\ref{eq:wavefunctionB}. In solid lines we plot Eq.~\ref{eq:CAaprox} choosing $q_c=\pi$.}
	\label{fig:4}
\end{figure}

One difference is that the MBS contribution at the middle-lower edge disappears abruptly for a critical $\Delta$. This behaviour appears in isotropic 3D band-edges~\cite{devega08a,navarretebenlloch11a,gonzaleztudela18d}, but not in 2D where the logarithmic divergence of $\Sigma_e$ at the band-edge guarantees the existence of the BSs for any $\Delta$~\cite{shi16a,calajo16a}. This is what happens in the other middle edge, where the BS contribution survives for $\Delta$'s inside of the band. Moreover, as the divergence at this band-edge scales as $1/\sqrt{x}$, the BS survives for a wider spectral region than the one appearing in the outer band-edges [see Fig.~\ref{fig:2}(c)]. Let us now quantitatively characterize these behaviours. On the lower band-edge $\Sigma_e(-\delta)=0$, such that the critical detuning where the MBS disappears in the lower band-edge is $\Delta_c=-\delta$. On the other band-edge, the BS survives even for $\Delta$'s close inside of the band due to the divergence of self-energy at this point. Asymptotically expanding $\Sigma_e(E)$ close to this band-edge, we estimate the MBS energy in the non-perturbative regime when $\Delta=\delta$, which scales as:
\begin{align}
 \label{eq:mbsenergy}
 E_\mathrm{MBS}-\delta\propto \sqrt[3]{\frac{g^4\delta}{J^3}\left[\log\left(\frac{J^4 C}{g^2\delta^2}\right)\right]^2}\,,
\end{align}
for $g^2\delta^2/J^2\ll 1$, where $C$ is a numerical factor. This formula agrees well with the results of solving exactly the pole equation [see Supp. Information]. For completeness, we also obtain the MBS wavefunction, which in the single-excitation regime has the general form:
\begin{align}
 \label{eq:wavefunction}
 \ket{\Psi}_\BS=\left[C_e\sigma_{eg}+\sum_{\kk}\left(C_{a}(\kk) a^\dagger_\kk +C_{b}(\kk) b_\kk^\dagger\right)\right]\ket{\mathrm{vac}}\,,
\end{align}
where $\ket{\mathrm{vac}}$ is the global vacuum of the combined QE-bath system. Solving $H \ket{\Psi}_\BS=E_\BS \ket{\Psi}_\BS$, one arrives to:
\begin{align}
 \label{eq:wavefunctionA}
 C_A(\kk)&\propto \frac{E_\BS+\cos(2\theta_\kk)\omega(\kk)}{E_\BS^2-\omega(\kk)^2}e^{-i\kk\nn_e}\,,\\
 C_B(\kk)&\propto -\frac{\omega(\kk)\sin(2\theta_\kk)}{E_\BS^2-\omega(\kk)^2}e^{-i\left(\kk\nn_e+\phi(\kk)\right)}\,,\label{eq:wavefunctionB}
\end{align}
for a QE coupled to the A lattice site at position $\nn_e$. The spatial distribution is obtained by Fourier transforming the previous expressions, which we plot in Figs.~\ref{fig:4}(a-b) for $E_\mathbf{BS}=0$, and a bath/gap sizes of $N=2^8$ and $\delta=0.01J$, respectively. The main feature is that the MBS has a very anisotropic wavefunction, being localized along the horizontal/vertical lines. This is reminiscent from the directional emission appearing at 2D Van Hove singularities. The underlying reason is that the $\kk$-points giving rise to the directional emission in the square geometry model are the ones that define the band-edge in this photonic superlattice. When breaking the lattice symmetry, we open a bandgap at these spectral region such that the bath excitation can not propagate and localize around the QE.

\begin{figure}[tb]
	\centering
	\includegraphics[width=0.85\linewidth]{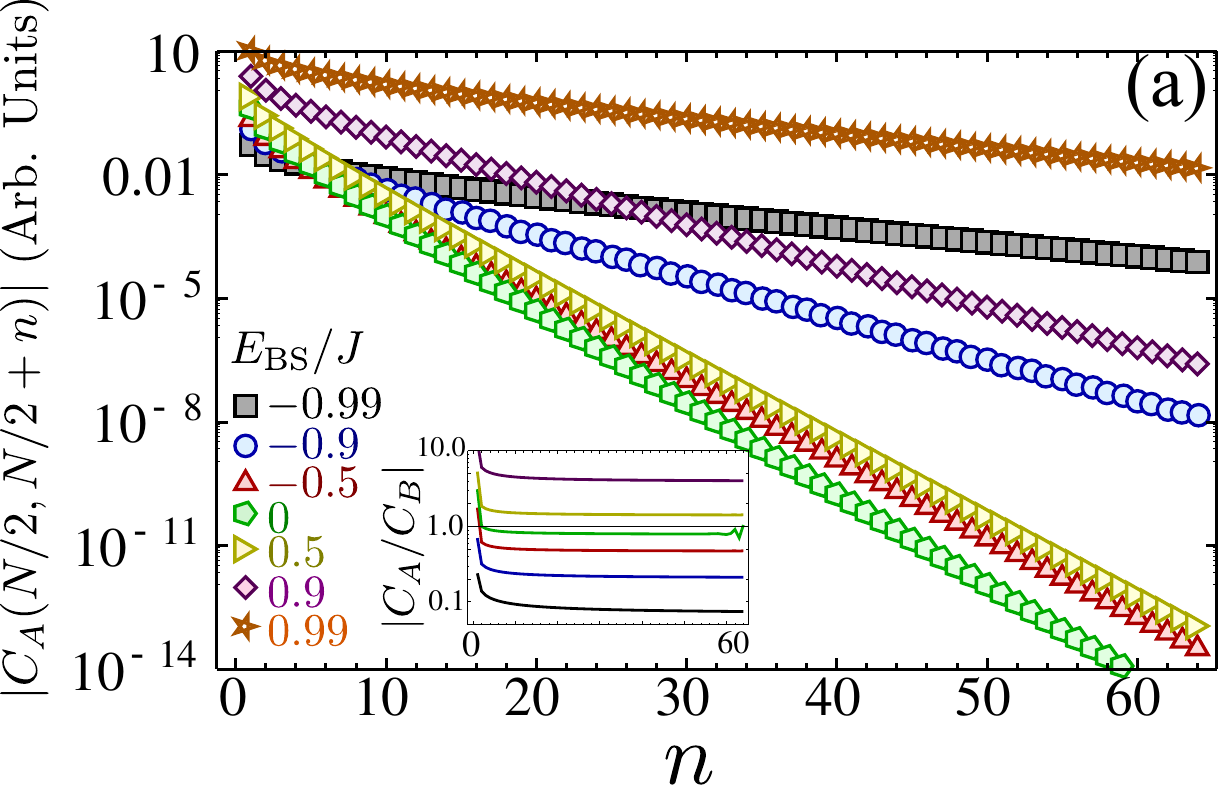}
	\caption{(a) Horizontal cuts of the $|C_{A}(n,0)|$, respectively, for a $N=2^8$ lattice with a fixed bandgap $\delta=J$ and different MBS energies as depicted in the legend. Inset: $|C_{A}(n,0)/C_{B}(n,0)|$ for the same parameters than panel (a).}
	\label{fig:5}
\end{figure}

In Figs~\ref{fig:4}(c-d), we make an horizontal cut of the bath BS wavefunction with energy $E_\BS=0$ at the A/B sublattices in the direction where the BS is less localized, and for several band-gap ratios $\delta/J$. For small $\delta/J$ we see that the wavefunction absolute value decays very slowly spreading almost along the entire bath size. We also observe that $|C_A|\gg |C_B|$ in this regime. For $\delta/J$ large the BS wavefunction starts to be exponentially localized around the impurity, and the A/B lattice starts to have similar decay behaviour (with small differences in distances closer to the QE). To qualitatively estimate the asymptotic decay of the wavefunction, e.g., $C_A(n,0)$, we realize that the larger contribution of the integrand of the Fourier Transform [$C_{A/B}(\kk)$] comes from the points closer to the band edge, $\omega(\kk)\approx \delta$, where $\cos(\theta_\kk)\approx 1$, $\sin(\theta_\kk)\approx 0$, and $\omega(\pi-q_1,\pi-q_2)\approx \left( \delta+\frac{q_1^2 q_2^2}{2\delta}\right)$. Using these expansions, we find that:

\begin{align}
\label{eq:CAaprox}
 C_A(n,0)\approx \frac{(-1)^n}{\pi\sqrt{2}}\Gamma(0,\sqrt{2}n\delta/q_c)\,.
\end{align}
where $\Gamma(a,z)$ is the incomplete $\Gamma$-function and $q_c=\pi$ a numerical cut-off we introduce to make the expressions converge [see Supp. Information]. In Figs.~\ref{fig:4}(c-d) we plot in solid lines Eq.~\ref{eq:CAaprox} to compare it with the numerical evaluation of the sums for a finite system. We observe that they show a qualitative agreement, even though it is not perfect for all $\delta/J$ and distances considered, possibly due to finite size effects and imprecise determination of $q_c$. One difference with respect to other reservoirs is that the BS length scales at the middle of the gap as $L_\BS\propto 1/\delta$, and not $1/\sqrt{\delta}$ like it typically occurs for bound states in all dimensions when $\omega(\kk)\propto |\kk|^2$. Furthermore, using the analytical expansions of the $\Gamma$-function we can obtain the approximated scaling of the wavefunction, yielding a logarithmic [Yukawa-type] spatial decay when $\sqrt{2}n\delta/q_c\ll [\gg] 1$, respectively.

For completeness, we study the situation where we fix the gap, e.g., $\delta=J$, and study the dependence of the wavefunctions as $E_\BS$ gets closer to the upper/lower band-edge. Let us summarize our findings: i) the wavefunctions display qualitatively the same anisotropic behaviour than in Fig.~\ref{fig:4}, with a change in the spatial distribution depending on $E_\BS$ [see Fig.~\ref{fig:5} to observe the change of $C_A$ along the main axis]; ii) the BS localizes mostly around A/B by getting closer to the upper/lower middle band-edge, respectively [see inset of Fig.~\ref{fig:5}]; the wavefunctions are lower bounded by the decay at $E_\BS=0$ (green pentagons), that is, the MBS length increases as $\Delta$ moves closer to the band-edge.

\section{Many QE: effective spin models\label{sec:many}}

\begin{figure}[tb]
	\centering
	\includegraphics[width=0.85\linewidth]{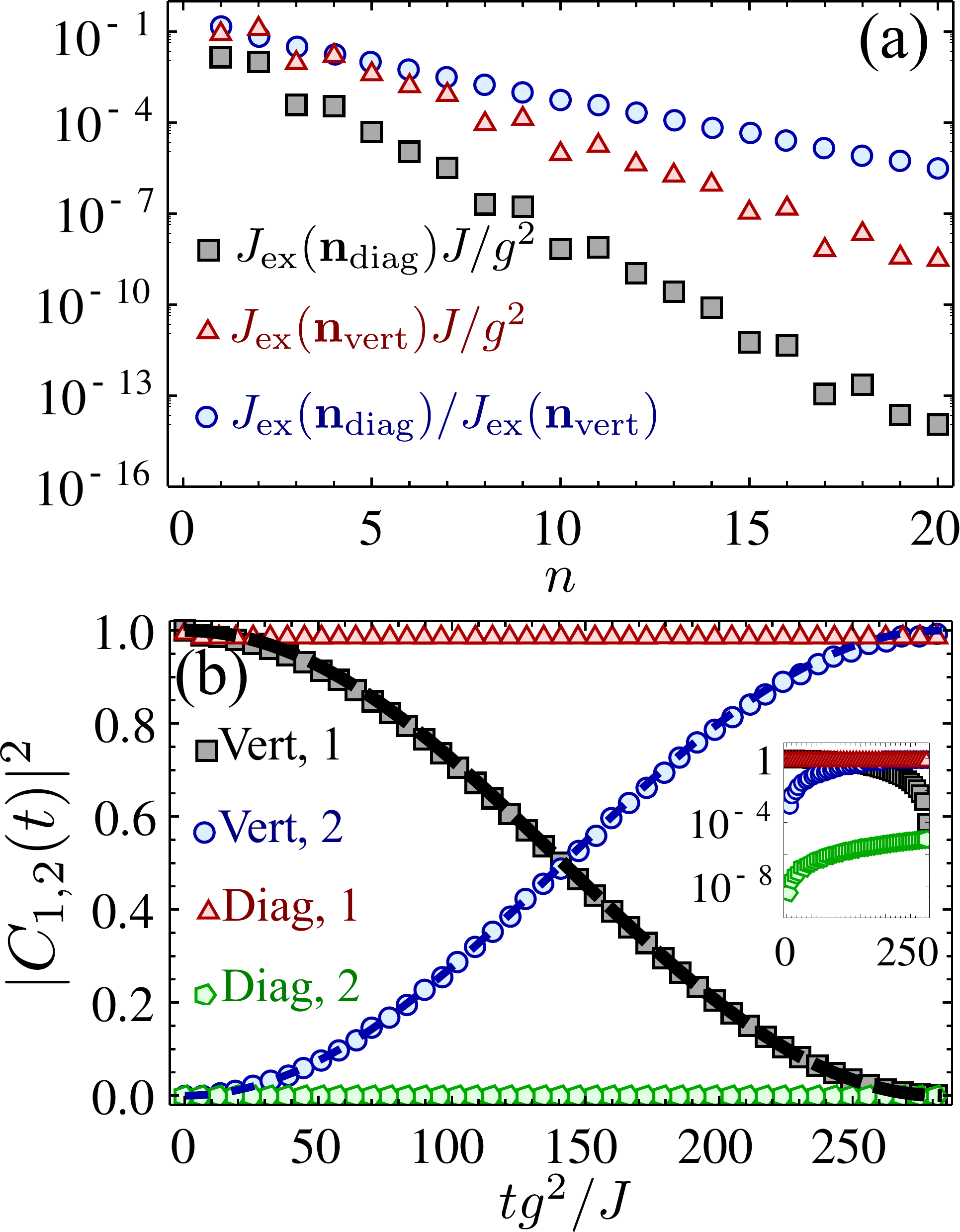}
	\caption{(a) Dipole-dipole coupling between two QEs coupled either vertically (red triangles), diagonally (black squares) and its ratio (blue spheres). (b) Two QE dynamics when one of them is initially excited for $n=6$ by solving the full QE-bath Hamiltonian. In black squares/blue markers (red triangles/green pentagons) we plot when the two QEs are placed vertically (diagonally). In dashed lines we plot $|\cos(J_\mathrm{ex} t)|^2/|\sin(J_\mathrm{ex} t)|^2$ with $J_\mathrm{ex}$ calculated from the exact pole equation. Inset: same figure but in logarithmic scale. Parameters: $N=2^8$, $g=0.1J$ and $\delta=J$. }
	\label{fig:6}
\end{figure}

In this section we explore the consequences of the MBS when multiple QEs interact with the bath. For concreteness, we assume all QEs couple to the same sublattice (A), and that their energies lie in the middle bandgap, where the dynamics is dominated by the MBS. We study two situations: i) one with two QEs in different positions to show that the interactions indeed behave as one expects from the shape of the MBS wavefunction; ii) as an application, we devise a protocol to entangle four distant qubits exploiting the highly anisotropic couplings emerging in these reservoirs and an extra auxiliary qubit.

\subsection{Two QEs: anisotropic dipole-dipole interactions}

Since our bath satisfies $\omega(-\kk)=\omega(\kk)$ and $\phi(\kk)=-\phi(-\kk)$, it can be shown~\cite{gonzaleztudela18c} that the dynamics of the symmetric/antisymmetric subspaces, $\sigma_{\pm}=(\sigma^1_{ge}\pm\sigma_{ge}^2)/\sqrt{2}$ decouple when two QEs are coupled to the bath. If they couple, e.g., to the A sublattice, the self-energy governing the symmetric/antisymmetric subspace reads:   $\Sigma^{\mathrm{AA}}_{\pm}(z;\nn_{12})=\Sigma_e^A(z)\pm \Sigma^{\mathrm{AA}}_{12}(z;\nn_{12})$, where:
\begin{align}
 \label{eq:selfplusmin}
 \Sigma^{\mathrm{AA}}_{12}(z;\nn_{12})=\frac{g^2}{N^2}\sum_{\kk} \frac{\mathrm{Re}\left[e^{i\kk\cdot\nn_{12}}\right]\cos(2\theta_\kk)\omega(\kk)}{z^2-\omega(\kk)^2}\,.
\end{align}

Here, it is clear that the distance dependence of the QE interactions is governed by the same expression as $C_A(\kk)$ in Eq.~\ref{eq:wavefunctionA}, so it inherits the same properties. To evidence this connection, we study the coherent transfer of excitations when one QE is initially excited, i.e.,  $\ket{\Psi(0)}=\ket{e_1,g_2}\otimes\ket{\mathrm{vac}}$. Since the dynamics between the symmetric and antisymmetric subspace decouples, we can solve the dynamics exactly~\cite{gonzaleztudela17b,gonzaleztudela18c} and find:
\begin{align}
\label{eq:dynexact}
 |C_{1 [2]}(t)|^2\approx R\cos^2(J_\mathrm{ex} t)\left[\sin^2(J_\mathrm{ex} t)\right]+\mathrm{others}\,, 
\end{align}
where $J_\mathrm{ex}=(E_{\BS,-}-E_{\BS,+})/2$ is the frequency of the oscillations obtained by solving the pole equations: $E_{\BS,\pm}=\Delta+\Sigma_{\pm}^{\mathrm{AA}}(E_{\BS,\pm};\nn_{12})$. The number $R(\le 1)$ is the residue of the symmetric/antisymmetric BS. There are other contributions which appear in the dynamics, e.g., branch-cut detours. However, here we focus on a parameter regime where $R\approx 1$, such that the rest of their contributions are small. Thus, the main effect of the interaction with the bath is a coherent exchange of excitations with frequency $J_\mathrm{ex}$.

In Fig.~\ref{fig:6} we compare the dynamics of two QEs placed vertically/diagonally, i.e., $\mathbf{n}_\mathrm{vert}/\mathbf{n}_\mathrm{diag}=(n,0)/(n,n)$, which we expect to interact strongly/weakly from the shape of the MBS wavefunction. In Fig.~\ref{fig:6}(a) we plot the expected dipole-dipole couplings as a function of $n$ by solving the exact pole equation in the symmetric/antisymmetric subspace. We see that both couplings are exponentially attenuated, as expected from Eq.~\ref{eq:CAaprox}, but the diagonal one decays exponentially faster than the vertical one [see blue spheres in Fig.~\ref{fig:6}(a)]. This different energy scales in the vertical/diagonal couplings have strong consequences on the dynamics, as we show in Fig.~\ref{fig:6}(b), where we plot the dynamics of the excited state of the two QEs.  There, we observe quasi-perfect state transfer between the QEs placed vertically, while the diagonal configuration shows almost no transfer within the timescales considered.

\subsection{Many QEs: long-distance entanglement trough auxiliary QEs}

Finally, we devise a simple application which exploits the shape of the emergent interactions. It consists in generating non-local entanglement among 4 qubits through an auxiliary one. At the end of the operation the auxiliary qubit is unentangled from the rest, so it leads to a pure entangled state among the four targeted qubits. 

\begin{figure}[!tb]
	\centering
	\includegraphics[width=0.99\linewidth]{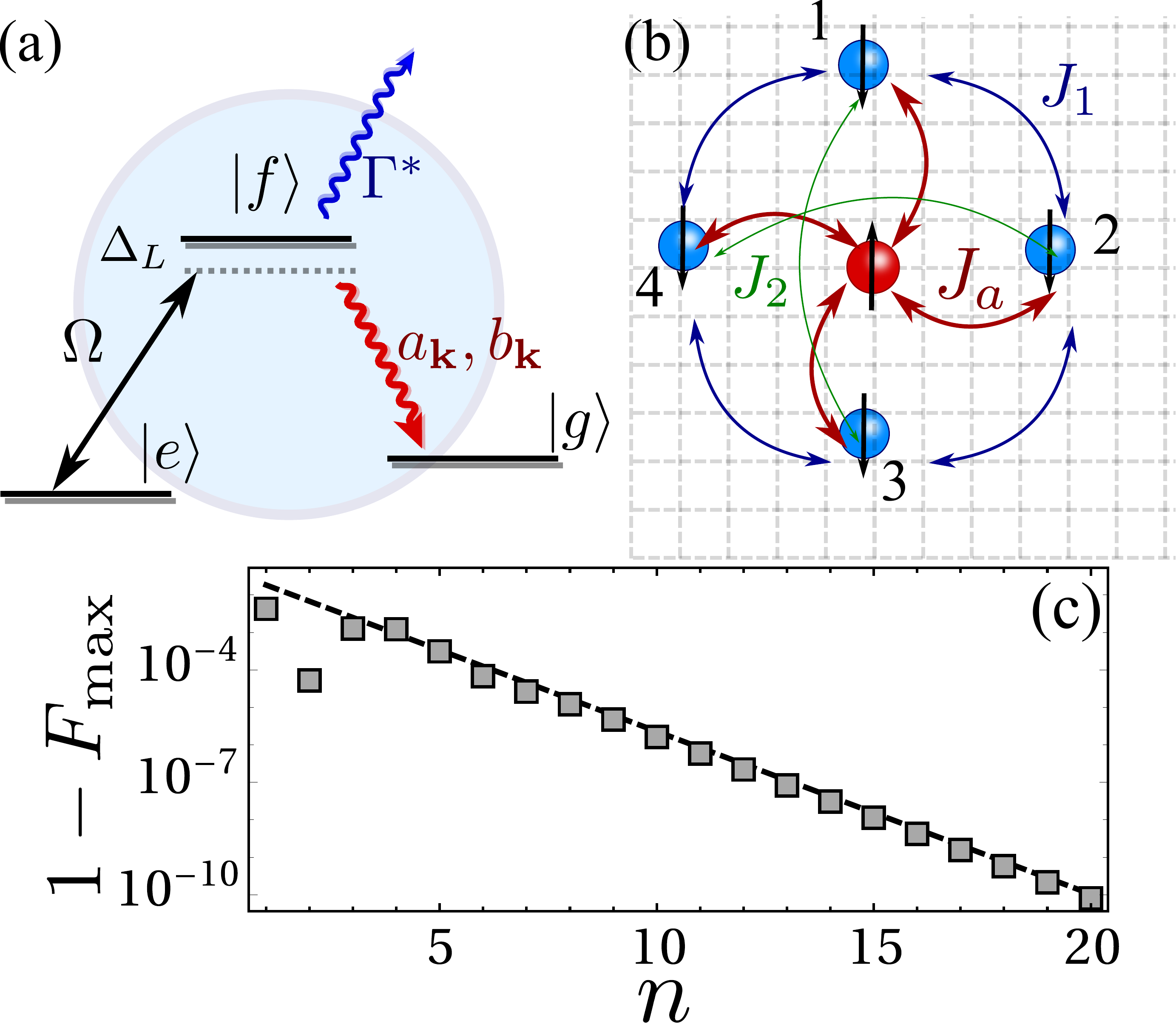}
	\caption{(a) Atomic $\Lambda$-scheme: one optical transition g-f couples to the bath, while the other one e-f couples with a classical field with amplitude $\Omega$ and detuning $\Delta_L$. (b) Scheme: 4 QEs distributed in a square with an auxiliary QE in the middle. (c) $1-F_\mathrm{max}$ (defined in Eq.~\ref{eq:maxfid}) as a function of the distance $n$ between the auxiliary QE and the rest. Parameters:  $g=0.1J$, $\delta=J$, $\Delta=0$ and $N=2^8$.}
	\label{fig:7}
\end{figure}

We define the qubits in the hyperfine levels ($g/e$) of an atomic $\Lambda$-scheme as depicted in Fig.~\ref{fig:7}(a). They are connected through an optically excited state, $f$, via a Raman laser in the $e-f$ transition, and through the bath in the $g-f$ transition. When the detuning of the Raman laser satisfies $|\Delta_L|\gg \Omega$, the excited state $f$ can be adiabatically eliminated~\cite{douglas15a,gonzaleztudela15c}, leading to an effective dynamics in the $\{g,e\}$ subspace described by the same Hamiltonian $H$, but with renormalized coupling rates and frequencies: $g\rightarrow g(\Omega/\Delta_L)$, $\Delta\rightarrow \Delta-\omega_L$. Both the bath induced timescales ($J_\mathrm{ex}$) and spontaneous emission ($\Gamma^*$) reduce by a factor $(\Omega/\Delta_L)^{-2}$, such that the ratio between coherent/incoherent processes remains unaltered. One advantage is that the QE-bath couplings can be switched dynamically with the Raman laser, and once the entanglement is created in the qubits is unaffected by free-space decay.

\begin{figure*}[!tb]
	\centering
	\includegraphics[width=0.9\linewidth]{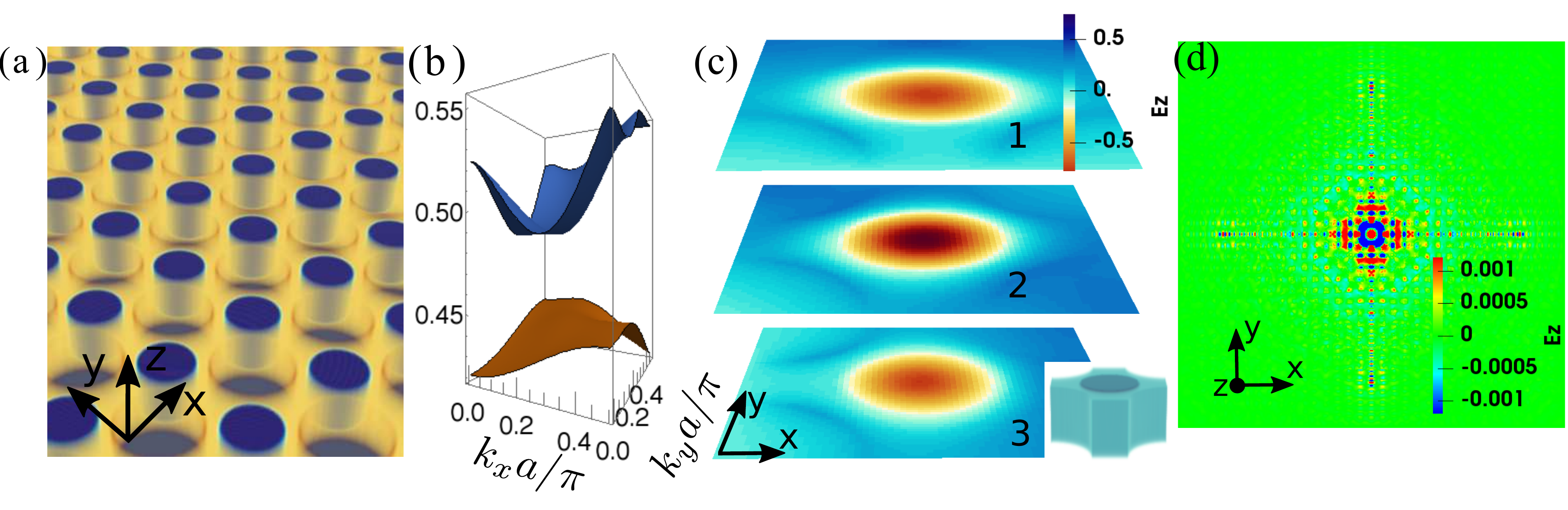}
	\caption{(a) Photonic crystal slab made of diamond with air/GaAs `holes'. Parameters: slab thickness $h=0.5a$, holes radii $r_1=0.25a$ (GaAs) and $r_2=0,35a$ (air) with $a$ the lattice spacing.  (b) Dispersion relation of the 3rd and 4th z-odd bands $\omega(k_x,k_y,0)$. (c) Electric field's spatial profile $E_z$ for the 3 momentum points $(k_xa/\pi,k_ya/\pi)=(0,0.27),(0.21,0.22),(0.27,0.01)$, at $z=0$ within the unit cell. Inset: dielectric function, with GaAs (dark blue) in the center of the cell, diamond (light blue) and air (white). The units are such that the field is normalized as $\int  \epsilon|{\bf E}|^2d^3{\bf x}=1$ in the unit cell. (d) FDTD simulation of the radiation from a dipole resonant with the upper (4th) z-odd band, of a crystal slab with a total of 34x34 unit cells. }
	\label{figCRYSTAL}
\end{figure*}

The protocol configuration is the following [see Fig.~\ref{fig:7}(b)]: four QEs are placed at $(\pm n, 0)$ and $(0,\pm n)$, and the auxiliary one (in red) at the origin. Since the distance between the four QEs to the auxiliary QE is the same, they all interact with it at a rate $J_a(n)$. Besides, the four QEs can talk directly between themselves either diagonally, with rate $J_1(n)$, or vertically/horizontally, $J_2(n)$. Thus, the effective Hamiltonian driving the evolution is:
\begin{align}
H_\mathrm{eff}&=J_a(n)\sigma^a_{eg}\sum_{i=1}^4 \sigma_{ge}^i+J_{1}(n)\left(\sigma_{eg}^1+\sigma_{eg}^3\right)\left(\sigma_{ge}^2+\sigma_{ge}^4\right) \nonumber \\ & + J_{2}(n)\left(\sigma_{eg}^1\sigma_{ge}^3+\sigma_{eg}^2\sigma_{ge}^4\right)+\hc\,\label{eq:Heff}
\end{align}

When all the Raman lasers act equally in all the QEs, and without considering other decoherence sources, the entangling protocol works as follows: First, we initialize the QEs in $g$. Next, we switch the auxiliary QE to the $e$ state with a microwave $\pi$-pulse, such that  $\ket{\Psi_1}=\ket{e}_a\otimes\ket{g}^{\otimes 4}$. Afterwards. we turn on all the Raman lasers, $\Omega$, such that the QEs can interact via the bath. Since initially there is a single excitation and $H_\mathrm{eff}$ conserves the number of excitations, it can be projected to the states: $\{\sigma_{eg}^{i}\ket{g}^{\otimes 5}\}$, with $i=a,1,2,3,4$. As we show in the Supp.~Information the probability to create the desired state, $\ket{\Psi}_\mathrm{goal}= \ket{g}_a\otimes \frac{1}{2}\sum_{i}\sigma_{eg}^i\ket{g}^{\otimes 4}$, is maximized if we keep the Raman lasers on for a time $\sqrt{(2 J_1+J_2)^2+16 J_a^2}T=\pi$, yielding a maximum fidelity of the entanglement generation:

\begin{align}
 F_\mathrm{max}&=|\bra{\Psi_\mathrm{goal}}e^{-i H_\mathrm{eff} T}\ket{\Psi_1}|^2=\frac{16 J_a^2}{(2 J_1+J_2)^2+16 J_a^2}\,.\label{eq:maxfid}
\end{align}

There, it is clear that the main limitation of the fidelity of the protocol is the cross-talk between the 4 target QEs, which provides an energy off-set between the levels, $\ket{\Psi}_\mathrm{goal}$ and $\ket{\Psi_1}$ preventing the complete excitation transfer. In Fig.~\ref{fig:7}(c) we show this error ($1-F_\mathrm{max}$) can be made small by increasing the distance $n$ between the QEs since $|(2J_1(n)+J_2(n))/J_a(n)|\ll 1$ for $n\gg 1$. Furthermore, if one allows for individual control of the Raman laser frequencies, we can impose $\Delta_a=\Delta+2J_1+J_2$ to correct the energy off-set of the cross-talk without imposing any restriction of the distances (as long as the different frequencies do not alter the effective spin interactions). Using this trick, one can achieve perfect coherent transfer by choosing the correct time duration of the Raman lasers, that is, $T=\pi/(4 J_a)$. The impact of free-space spontaneous emission from the QEs can be accounted in our protocol by replacing $H_\mathrm{eff}$ with a non-unitary Hamiltonian: $H_\mathrm{eff}^*=H_\mathrm{eff}-i\frac{\Gamma^*}{2}\sum_{j}\sigma_{ee}^j$. With it, the maximal fidelity of the protocol can be calculated as in Eq.~\ref{eq:maxfid}, arriving to $F_{\mathrm{max},*}=e^{-\Gamma^*\pi/(2 J_a)}$. Other error sources can be considered in a similar fashion.

\section{Photonic crystal realization \label{sec:bath}}

Until now, we have used a simple model of a photonic crystal which allows us to extract mathematical and physical intuition of the phenomena that we want to explore. To complete the manuscript, we give a proof-of-principle example on how to obtain the desired band structure with a real photonic crystal implementation. The challenge consists in finding a geometry which both opens a robust bandgap at the original Van-Hove singularity, and where other bands do not enter in the middle. 

One configuration we found consists of a dielectric slab with $\epsilon_r=6$, such as diamond, with two periodic lattices of circular defects, the larger ones empty, while the other ones filled with a larger index dielectric $\epsilon_r=13$, like GaAs [see Fig.~\ref{figCRYSTAL}(a)].  As shown in Fig.~\ref{figCRYSTAL}(b), a suitable middle-band gap appears in our configuration between the third and fourth band of z-odd modes (i.e. TM 
at $z=0$). In Fig.~\ref{figCRYSTAL}(c), we show the electric field, $E_z$, associated to several representative $\kk$-points (see caption), to show which are the most appropriate regions to place our QE. The air gaps, suited for optically trapping atoms~\cite{thompson13a,goban13a,hood16a}, and the GaAs cylinders, where one could implant solid state emitters~\cite{lodahl15a,sipahigil16a}, are the regions with higher field concentration. In addition, by symmetry, the plane $z=0$ has maximum value of $E_z$ for z-odd modes and nearly vanishing (not shown) for z-even (TE in this plane). In this way, a dipole aligned to the $z$ direction will couple mostly to z-odd modes, as desired. 

Compared to the simplified model considered in the manuscript, the two bands are not symmetric with respect to the middle of the gap, so the behaviour will be different from the predicted one~\footnote{Actually, only the upper (4th) z-odd band is suitable to mediate anisotropic interactions previously described.}. To estimate to what extent the anisotropic behaviour survives, we do a finite-difference time-domain simulation~\cite{MEEP} of the emission of an atomic dipole aligned along $z$ and resonant with the 4th band. In Fig.~\ref{figCRYSTAL}(c), we observe that the emission still occurs in a highly directional fashion, which foresees the observation of QE anisotropic dipole interactions when the atomic frequencies lie in the band-gap. Even though this configuration might not be the optimal one, it provides a first direction on how to engineer these band-gaps with photonic crystals. Further optimization can be obtained with, e.g., inverse design~\cite{sapra18a}, although this lies beyond the scope of the manuscript.

\section{Conclusions \label{sec:conclu}}

Summing up, we have shown how the directional collective dissipation appearing in 2D Van-Hove singularities turns into anisotropic coherent interactions by using photon superlattices. The key point is the emergence of an extra bound state which inherits the directional properties of the emission. Using a simple description, we characterize the different dynamical regimes appearing, the anisotropic bound state properties, and the effective spin-models when many emitters interact with the bath. Apart from the photonic-crystal implementations in the optical regime, of which we provide a proof-of-principle example, there are other platforms where our findings are extrapolable, such as circuit QED systems~\cite{liu17a,sundaresan18a,mirhosseini18a,Houck2018} or cold-atoms in state-dependent optical lattices~\cite{devega08a,navarretebenlloch11a,krinner18a}.

Beyond the applications highlighted along the manuscript, like the controlled generation of long-range entanglement, other interesting directions to explore are self-organization phenomena under these very anisotropic potentials~\cite{chang13a}, design repulsive fermionic potentials to simulate pseudo-molecules ~\cite{luengo18a}, or the study of the multi-photon behaviour of these directional bound states~\cite{shi18a}.

\section*{Acknowledgements}
AGT acknowledges the ERC Advanced Grant QENOCOBA under the EU Horizon 2020 program (grant agreement 742102).  FG acknowledges MINECO/AEI/FEDER through projects QuStruct FIS2015-66860-P and EPheQuCS FIS2016-78010-P. We also acknowledge discussions with J.~Arg\"uello-Luengo and J. I.~Cirac.\\

\pagebreak
\newpage

\begin{widetext}
	\begin{center}
\textbf{\large Supplementary Information: Anisotropic quantum emitter interactions in two-dimensional photonic-crystal baths.}
\end{center}

\setcounter{equation}{0}
\setcounter{figure}{0}
\setcounter{section}{0}
\makeatletter

\renewcommand{\thefigure}{SM\arabic{figure}}
\renewcommand{\thesection}{SM\arabic{section}}  
\renewcommand{\theequation}{SM\arabic{equation}}

In this Supplementary Material, we provide more details on: i) Asymptotic expansions of the single QE self-energy; ii) Analytical continuation of the single quantum emitter (QE) self-energy to perform the exact integration of the excited state dynamics, $C_e(t)$; iii) Asymptotic scaling of the spatial decay of the extra bound state appearing in the middle band-gap; iv) Calculation of the fidelity of the four-qubit entangling protocol.

\section{Self-energy expansions and analytical approximation of middle bound state energies.}

The analytical formula for the self-energy for a QE coupled to the A sublattice reads:
\begin{align}
\Sigma_{e}(z)=\frac{2 g^2 (z+ \delta)}{\pi (z^2-\delta^2)}\KK\left[\frac{16 J^2}{z^2-\delta^2}\right]\,.\label{eq:SManalself}
\end{align}

From here, it can be shown that around the middle band-edges:
\begin{align}
\label{eq:scaling}
\Sigma_e(\delta+x+ i 0^+)\approx \frac{g^2}{\sqrt{8}}\sqrt{\frac{\delta}{x}}\left[1+\frac{i}{\pi}\log\left(\frac{\delta x}{128 J^2}\right)\right]\,,\\
\Sigma_e(-\delta-x+ i 0^+)\approx \frac{g^2}{\sqrt{32}}\sqrt{\frac{x}{\delta}}\left[-1+\frac{i}{\pi}\log\left(\frac{\delta x}{128 J^2}\right)\right]\,. 
\end{align}
for $0<x\ll J$. Thus, one middle band edge diverges as $1/\sqrt{x}$, as in 1D reservoirs, while the other one goes to zero as $\sqrt{x}$ as is the case in isotropic 3D reservoirs. 

We can use these expansions, for example, to obtain an analytical approximation to the existence conditions of the middle-bound state (MBS), which is the main focus of the manuscript. On the lower band-edge $\Sigma_e(-\delta)=0$, such that the critical detuning where the MBS disappears in the lower band-edge is just $\Delta_c=-\delta$. On the other band-edge, we can expand the self-energy for energies below the gap, finding:
\begin{align}
\label{eq:scaling2}
\Sigma_e(\delta-x+ i 0^+)\approx -\frac{g^2}{\sqrt{8}\pi J}\sqrt{\frac{\delta}{x}}\log\left(\frac{128 J^2}{\delta x}\right)\,,
\end{align}
for $x\ll J$. Using this expansion, we can solve the pole equation analytically to obtain the energy of the MBS when $\Delta=\delta$, yielding to:
\begin{align}
\label{eq:mbsenergy}
E_\mathrm{MBS}=\delta-\sqrt[3]{\frac{g^4 \delta}{2(3\pi)^2 J^3 }\left(W\left(\frac{6144\pi J^4}{g^2\delta^2}\right)\right)^2}\,,
\end{align}
which agrees very well with the results of numerically solving the pole equation, as shown in Fig.~\ref{fig:3}.

\begin{figure}[tb]
	\centering
	\includegraphics[width=0.3995\linewidth]{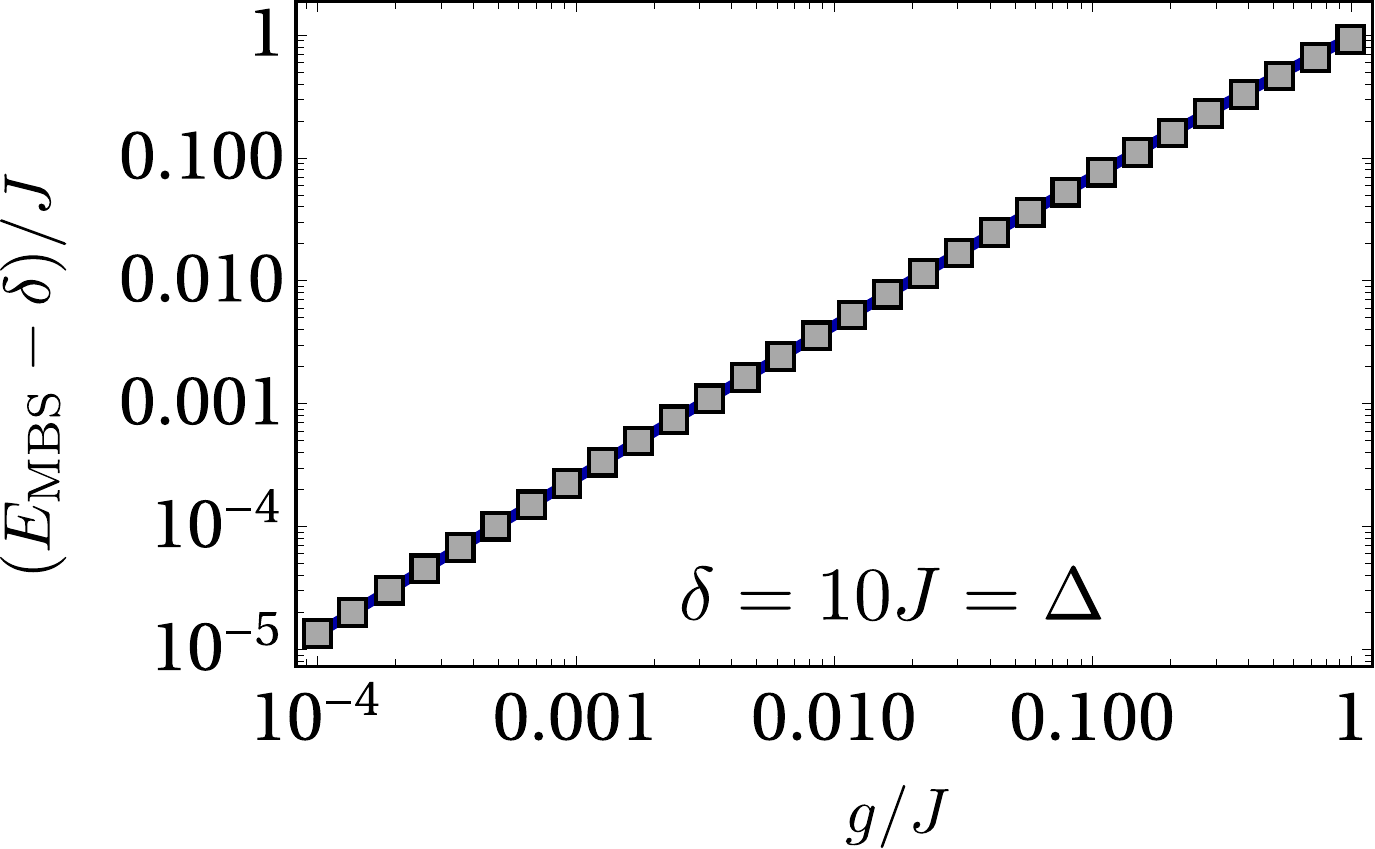}
	\caption{(a) $E_\mathrm{MBS}$ as a function of $g/J$ for a QE coupled to the A sublattice with $\Delta=\delta=10J$. In markers the numerical results obtained by solving the exact pole equation, while in solid line the analytical result of Eq.~\ref{eq:mbsenergy} obtained by expanding the solution of the self-energy close to the upper middle edge.}
	\label{fig:3}
\end{figure}

\section{Analytical continuation of the single QE self-energy}

To fully characterize the different dynamical regimes one must integrate Eq.~\ref{eq:ex} of the main text exactly. To do the exact integration, we transform the integral above the real axis into a complex one by closing the contour of integration to apply Residue Theorem. Since the self-energy $\Sigma_e(z)$ is not analytical in certain regions, one can not simply close the contour with a semiarc in the lower half-plane. One possible choice of the detours to avoid the non-analytical regions is to take $E_\BC\pm 0^\pm -i y$ for four values of $E_\BC=-\sqrt{16 J^2+\delta^2},-\delta,\delta,\sqrt{16 J^2+\delta^2}$.  This divides the lower complex plane in five regions, where the 
definition of $\Sigma_e(z)$ must be adapted. To guarantee that we go to the correct Riemann surface of $\Sigma_e(z)$:
\begin{itemize}
 \item In regions I, III, and V, defined by $\mathrm{Re}(z)\in (-\infty,-\sqrt{16J^2+\delta^2})$, $(-\delta,\delta)$ and $(\sqrt{16J^2+\delta^2},\infty)$, respectively, one must use the definition of $\Sigma_e(z)$ as written in Eq.~\ref{eq:analself} of the main text.
 
 \item In regions II and IV, defined by $\mathrm{Re}(z)\in (-\sqrt{16J^2+\delta^2},-\delta)$, and $(\delta,\sqrt{16J^2+\delta^2})$, one must adapt the definition of the self-energy as follows:
\begin{align}
 \Sigma_{e}(z)=\frac{2 g^2 (z+ \delta)}{\pi^2 (z^2-\delta^2)}\left(\KK\left[\frac{16 J^2}{z^2-\delta^2}\right]\pm 2i\KK\left[1-\frac{16 J^2}{z^2-\delta^2}\right]\right)\,,\label{eq:analself2}
\end{align}
\end{itemize}

With these changes in the definition, one can now perform the exact integration of the dynamics separating the different contributions, as we did in Fig.~\ref{fig:2} of the main text.

\section{Asymptotic scaling of the BS wavefunction}

The bound state wavefunction of a single QE in the single excitation subspace for our type of bath generally has the form:
\begin{align}
 \label{eqSM:wavefunction}
 \ket{\Psi}_\BS=\left[C_e\sigma_{eg}+\sum_{\kk}\left(C_{a}(\kk) a^\dagger_\kk +C_{b}(\kk) b_\kk^\dagger\right)\right]\ket{\mathrm{vac}}\,,
\end{align}
where $\ket{\mathrm{vac}}$ is the global vacuum of the combined QE-bath system. By solving $H \ket{\Psi}_\BS=E_\BS \ket{\Psi}_\BS$, one arrives to:
\begin{align}
 \label{eqSM:wavefunctionA}
 C_A(\kk)&\propto \frac{E_\BS+\cos(2\theta_\kk)\omega(\kk)}{E_\BS^2-\omega(\kk)^2}e^{-i\kk\nn_e}\,,\\
 C_B(\kk)&\propto -\frac{\omega(\kk)\sin(2\theta_\kk)}{E_\BS^2-\omega(\kk)^2}e^{-i\left(\kk\nn_e+\phi(\kk)\right)}\,.\label{eqSM:wavefunctionB}\,,
\end{align}
for a QE coupled to the A lattice site at position $\nn_e$. To make a more quantitative estimation of the decay of the wavefunction, we can consider that the larger contribution to the integrand of $C_{A/B}(\kk)$ will come from the points closer to the band edge, $\omega(\kk)\approx \delta$. At these points $\cos(\theta_\kk)\approx 1$, $\sin(\theta_\kk)\approx 0$, and the energy dispersion is expanded, e.g., 
\begin{equation}
 \omega(\pi-q_1,\pi-q_2)\approx \delta\left(1+\frac{q_1^2 q_2^2}{2\delta^2}\right)\,,
\end{equation}
for $q_{1,2}\ll 1$. Focusing on $C_A(\nn)$, the sum of the contribution around the band-edge frequencies can be rewritten:
\begin{align}
 C_A(\nn)\approx \frac{2}{(2\pi)^2\delta}\Bigg[&(-1)^{(n_1+n_2)}\iint_0^{2\pi}d^2\qq \frac{\mathrm{Re}\left[e^{i(q_1 n_1+q_2 n_2)}\right]}{1+\frac{q_1^2 q_2^2}{2\delta^2}}+\nonumber\\
 &(-1)^{(n_1-n_2)}\iint_0^{2\pi}d^2\qq \frac{\mathrm{Re}\left[e^{i(q_1 n_1-q_2 n_2)}\right]}{1+\frac{q_1^2 q_2^2}{2\delta^2}} \Bigg]\,.
\end{align}

To continue the derivation, let us restrict to a particular direction, e.g., $n_1\equiv n$ and $n_2=0$. Notice, that since we have made the expansion for $q_{i}\ll 1$, the upper limit of the integral should not matter since the main contribution will be coming from $q_i\rightarrow 0$. Then, we can in principle extend the integral to infinite. Using this, and the fact that:
\begin{align}
 \label{eq:auxint}
 \int_{0}^\infty dq_1 \frac{\cos(q_1 n)}{1+\frac{q_1^2 q_2^2}{2\delta^2}}=\frac{e^{-\sqrt{2}n\delta/q_2}\pi\delta}{\sqrt{2}q_2}\,,
\end{align}
we arrive to:
\begin{align}
 C_A(n,0)\approx \frac{(-1)^n}{\pi\sqrt{2}}\Bigg[\int_0^{\infty}dq_2 \frac{e^{-\sqrt{2}n\delta/q_2}}{q_2} \Bigg]\,.
\end{align}

The problem of the previous integral is that it does not converge because the integrand scales as $1/q_2$ when $q_2\rightarrow \infty$. However, we know there should be a natural cut-off given by discretization. Thus, we replace $\infty$ by $q_c$ and find that:
\begin{align}
\label{eq:CAaprox2}
 C_A(n,0)\approx \frac{(-1)^n}{\pi\sqrt{2}}\Gamma(0,\sqrt{2}n\delta/q_c)\,.
\end{align}
where $\Gamma(a,z)$ being the incomplete $\Gamma$-function. We can use the analytical expansions of the $\Gamma$ to obtain the approximated analytical scaling in the small/large band-gap limit, that is:
\begin{align}
\Gamma(0,x\ll 1)\approx -\gamma+\log(1/x)\,,\\
\Gamma(0,x\gg 1)\approx \frac{e^{-x}}{x}\,,
\end{align}
where $\gamma\approx 0.577$ is the Euler constant. Thus, the wavefunction shows a very slow logarithmic decay when $\sqrt{2}n\delta/q_c\ll 1$, while having a Yukawa-type decay when $\sqrt{2}n\delta/q_c\gg 1$.

\section{Four-qubit entangling protocol}

Let us study in detail how the entangling protocol works in the simplest configuration, that is, when all the Raman lasers act equally in all the atoms and without considering other decoherence sources. We assume that all the atoms are initially in the ground state, $\ket{\Psi_0}=\ket{g}_a\otimes\ket{g}^{\otimes 4}$, while the bath is also in the vacuum state. Next, with a microwave field we switch the auxiliary atom to the $e$ state with a $\pi$-pulse, such that  $\ket{\Psi_1}=\ket{e}_a\otimes\ket{g}^{\otimes 4}$. Then, we switch on all the Raman lasers, $\Omega$, such that the QEs can interact between themselves by exchanging/absorbing bath excitation with the assistance of the Raman laser. Since, we start effectively with a single excitation in the five QEs, the effective dynamics can be written in a subspace: $B=\{\ket{e}_a\otimes \ket{g}^{\otimes 4},\ket{g}_a\otimes \sigma_{eg}^{i}\ket{g}
^{\otimes 4}\}$, with $i=1,2,3,4$. In this basis, the effective Hamiltonian governing the interaction:
\begin{align}
 H_\mathrm{eff}=\left( {\begin{array}{ccccc}
   \Delta & J_a & J_a & J_a & J_a\\
   J_a & \Delta & J_1 & J_2 & J_1\\
   J_a & J_1 & \Delta & J_1 & J_2\\
   J_a & J_2 & J_1 & \Delta & J_1\\
  J_a & J_1 & J_2 & J_1 & \Delta\\
  \end{array} } \right)\label{eq:hameff}
\end{align}

It is instructive to rewrite this effective Hamiltonian in a basis, $B'=\{\ket{\alpha_i}\}_{i=1}^5$, that contains the state we want to obtain, that is: $\ket{\Psi}_\mathrm{goal}= \ket{g}_a\otimes \frac{1}{2}\sum_{i}\sigma_{eg}^i\ket{g}^{\otimes 4}$. This is achieved with the following unitary transformation:
\begin{align}
 U=\left( {\begin{array}{ccccc}
   1 & 0 & 0 & 0 & 0\\
   0 & \frac{1}{2} & \frac{1}{2} & \frac{1}{2} & \frac{1}{2}\\
   0 & \frac{1}{2} & \frac{1}{2} & -\frac{1}{2} & -\frac{1}{2}\\
   0 & \frac{1}{\sqrt{2}} &  -\frac{1}{\sqrt{2}} & 0 & 0\\
  0 & 0 & 0 &  \frac{1}{\sqrt{2}} & -\frac{1}{\sqrt{2}}\\
  \end{array} } \right)\,
\end{align}
where $\ket{\alpha_1}=\ket{\Psi_1}$ and $\ket{\alpha_2}=\ket{\Psi}_\mathrm{goal}$. In this new basis, the effective Hamiltonian reads:
\begin{align}
 H_\mathrm{eff}=\Delta\mathbf{1}+\left( {\begin{array}{ccccc}
   0 & 2 J_a & 0 & 0 & 0\\
   2 J_a & 2J_1+J_2 & 0 &0 & 0\\
   0 & 0 & -J_2 & 0 & 0\\
   0 & 0 & 0 & -J_1 & J_2-J_1\\
  0 & 0 & 0 & J_2-J_1 & -J_1\\
  \end{array} } \right)\,.
\end{align}

One immediately realizes that our initial state is indeed only coupled to the desired state, $\ket{\Psi}_\mathrm{goal}$, due to the spatial symmetry of our initial state. Then, the fidelity of the protocol can be obtained by solving the dynamics in the restricted $2\times 2$ subspace, yielding:
\begin{align}
 F&=|\bra{\Psi_\mathrm{goal}}e^{-i H_\mathrm{eff} t}\ket{\Psi_1}|^2=\frac{16 J_a^2}{R^2}\sin^2(R t/2)\,,
\end{align}
with $R=\sqrt{(2 J_1+J_2)^2+16 J_a^2}$. Thus, choosing the time duration of the operation $R T=\pi$, we maximize the excitation transfer and the fidelity, yielding:
\begin{align}
 F_\mathrm{max}&=|\bra{\Psi_\mathrm{goal}}e^{-i H_\mathrm{eff} t}\ket{\Psi_1}|^2=\frac{16 J_a^2}{(2 J_1+J_2)^2+16 J_a^2}\,.\label{eqSM:maxfid}
\end{align}

\end{widetext}
\bibliography{Sci,books}

\end{document}